\documentclass[aps,prl,twocolumn,,superscriptaddress,floatfix]{revtex4-1}
\usepackage[svgnames]{xcolor}
\usepackage[colorlinks=true,urlcolor = purple,linkcolor=teal,citecolor=magenta,bookmarks=false]{hyperref}
\usepackage{amsfonts,amsmath,amssymb}
\usepackage{graphicx}
\usepackage{dsfont}
\usepackage{graphicx}
\usepackage{amssymb}
\usepackage{amsmath}
\usepackage{empheq}

\usepackage{bm}

\newcommand{\red}{}

\newcommand{\bc}{\begin{center}}
\newcommand{\ec}{\end{center}}
\def\ba#1{\begin{array}{#1}\displaystyle}
\newcommand{\ea}{\end{array}}

\newcommand{\beq}{\begin{equation}}
\newcommand{\eeq}{\end{equation}}
\newcommand{\beqa}{\begin{eqnarray}}
\newcommand{\eeqa}{\end{eqnarray}}
\newcommand{\no}{\nonumber}
\newcommand{\n}{\nonumber\\}
\newcommand{\bi}{\begin{itemize}}
\newcommand{\ei}{\end{itemize}}

\def\lt#1{\left#1}
\def\rt#1{\right#1}

\def\frc#1#2{\frac{#1}{#2}}

\newcommand{\p}{\partial}

\newcommand{\bra}{\langle}
\newcommand{\ket}{\rangle}

\newcommand{\R}{{\mathbb{R}}}

\newcommand{\Or}{{\cal O}}

\newcommand{\ep}{\varepsilon}

\newcommand{\ri}{{\rm i}}
\newcommand{\dd}{{\rm d}}

\begin{document}
\title{Hydrodynamic Diffusion in Integrable Systems}

\author{Jacopo De Nardis}
\affiliation{D\'epartement de Physique, Ecole Normale Sup\'erieure,
PSL Research University, CNRS, 24 rue Lhomond, 75005 Paris, France}
\author{Denis Bernard}
\affiliation{Laboratoire de Physique Th\'eorique de l'Ecole Normale Sup\'erieure de Paris, CNRS, ENS, PSL University \& Sorbonne Universit\'e, France.}
\author{Benjamin Doyon}
\affiliation
{Department of Mathematics, King's College London, Strand WC2R 2LS, London, U.K.}

\date{\today}

\begin{abstract} We show that hydrodynamic diffusion is generically present in many-body interacting integrable models. We extend the recently developed generalised hydrodynamic (GHD) to include terms of Navier-Stokes type which lead to positive entropy production and diffusive relaxation mechanisms. These terms provide the subleading diffusive corrections to Euler-scale GHD for the large-scale non-equilibrium dynamics of integrable systems, and arise due to  two-body scatterings among quasiparticles. We give exact expressions for the diffusion coefficients. Our results apply to a large class of  integrable models, including quantum and classical, Galilean and relativistic field theories, chains and gases in one dimension, such as the Lieb-Liniger model describing cold atom gases and the Heisenberg quantum spin chain. We provide numerical evaluations in the Heisenberg spin chain, both for the spin diffusion constant, and for the diffusive effects during the melting of a small domain wall of spins, finding excellent agreement with tDMRG numerical simulations.
\end{abstract}

\maketitle

{\em Introduction.---} 
The study of quantum systems far from equilibrium has received a large amount of attention in recent years. In the context of cold atom gases, unprecedented experimental control means that it is now possible to observe the behavior of many-body quantum systems in fully inhomogeneous and isolated setups \cite{Hild14,Boll2016,1805.10990,1712.04642,1707.07031,Rauer2018}. The foremost theory for describing inhomogeneous many-body systems is hydrodynamics. Despite its long history, it has now seen a resurgence of interest, as it emerges in new contexts and finds new applications especially in the field of strongly correlated systems \cite{Bhaseen:2015aa,1511.03646,PhysRevLett.100.191601,PhysRevE.85.046408,Lucas2016,PhysRevD.95.096003}. Of particular interest in the present letter is the recent development of the hydrodynamic theory appropriate to integrable systems, dubbed generalised hydrodynamics (GHD) \cite{PhysRevX.6.041065,SciPostPhys.2.2.014,PhysRevLett.117.207201,PhysRevLett.119.220604,PhysRevB.97.045407,Bulchandani2017}. This theory applies to a large family of models, including the paradigmatic Heisenberg quantum chain for one-dimensional magnetism, see \cite{PhysRevLett.117.207201,PhysRevB.96.115124,PhysRevB.96.020403,PhysRevB.97.045407}, as well as the Bose gas with delta-function interaction (the Lieb-Liniger (LL) model) \cite{PhysRev.130.1605} which describes gases of ultracold atoms confined to one-dimensional traps, see \cite{1711.00873,PhysRevLett.119.195301}. GHD has given rise to a raft of new exact, sometimes unexpected, results in the past few years: it provides exact descriptions of steady states fully out of equilibrium \cite{PhysRevX.6.041065,PhysRevLett.117.207201}, it efficiently describes the famous quantum Newton cradle setup \cite{Kinoshita:2006aa,1707.07031} where lack of thermalisation is explicitly observed \cite{1711.00873,PhysRevLett.120.164101}, it characterizes the transport of quantum entropy through a chain \cite{1805.01884,1706.00020} and it gives exact expressions for Drude weights  \cite{IN_Drude,SciPostPhys.3.6.039,IN_Hubbard,PhysRevB.97.045407} characterising ballistic transport in integrable models \cite{Zotos99,PhysRevB.55.11029}.

The dynamical observables in hydrodynamic theories are the conserved densities afforded by the microscopic dynamics. Let $Q_i = \int_\R \dd x\,\mathfrak{q}_i(x)$ be conserved charges, which commute with the Hamiltonian and with each other, and $\mathfrak{q}_i(x,t)$ and $\mathfrak{j}_i(x,t)$ their charge and current densities,
\beq\label{conslaw}
	\p_t \mathfrak{q}_i(x,t) + \p_x \mathfrak{j}_i(x,t) = 0.
\eeq
Hydrodynamics assumes weak time and space modulations of expectation values of charge and current densities, and is formulated as a theory for the space-time evolution of their averages,
\beq\label{ghdcons}
	\p_t \bra\mathfrak{q}_i(x,t)\ket + \p_x\bra\mathfrak{j}_i(x,t)\ket = 0.
\eeq
The averages $\bra\mathfrak{q}_i(x,t)\ket$ are seen as independent local-state variables, and the currents $\bra\mathfrak{j}_i(x,t)\ket$ as functions of these state variables and their space derivatives,
\beq\label{constitutive}
	\bra\mathfrak{j}_i(x,t)\ket = {\cal F}_i(x,t) + \sum_j {\cal F}_{ij}(x,t)\p_x\bra\mathfrak{q}_j(x,t)\ket + \ldots
\eeq
The quantities ${\cal F}_i(x,t)$, ${\cal F}_{ij}(x,t)$, $\ldots$ are functions of all averages $\bra\mathfrak{q}_k(x,t)\ket$ at the point $x,t$, whence their space-time dependence, and encode the model properties.

The terms ${\cal F}_i(x,t)$ represent the \emph{Euler scale} hydrodynamic theory \cite{Spohn1991}.  The assumption is that all local fluid cells are, to leading approximation, at thermal equilibrium. This means that, locally, the reduced density matrix is stationary and it is then given by a (generalised) Gibbs ensemble (GGE), proportional to (see the reviews \cite{reviewGGE1,Essler:2016aa,1742-5468-2016-6-064007})
\beq\label{GGE}
	e^{-\sum_i \beta_i(x,t) Q_i}.
\eeq
The Lagrange parameters $\beta_i(x,t)$ are fixed by the average values of the densities, $\bra \mathfrak{q}_i(x,t)\ket$, and in turn determine the currents $\bra\mathfrak{j}_i(x,t)\ket$: these are the equations of state. Euler hydrodynamics is time-reversible, and describes ballistic transport. It applies when variations in space and time occur at very large scales only.
However, Euler hydrodynamics often develops instabilities such as large gradients and fails to describe the loss of large-scale structures over time \cite{Bressan2013,Spohn1991}, see \cite{PhysRevB.96.115124,Misguich17,Ljubotina_nature} for recent observations in integrable systems.

Beyond the Euler scale, the terms involving spatial derivatives cannot be determined by the homogeneous, stationary thermodynamics of the gas. They are referred to as {\em constitutive relations} of the hydrodynamic theory, and must be fixed in an alternative fashion. Often, one takes into account various symmetries available and fixes them in a phenomenological fashion. Of particular importance are the terms of Navier-Stokes type, $\sum_j {\cal F}_{ij}(x,t)\p_x\bra\mathfrak{q}_j(x,t)\ket$: they give rise to diffusive effects, the irreversible processes by which large-scale structures are passed to mesoscopic-scale fluid cells and increase their entropy.

Euler hydrodynamics is very relevant to integrable models, as the infinite number of conserved quantities guarantees a large amount of ballistic transport. In integrable models, the homogeneous steady states \eqref{GGE} involve infinitely many Lagrange parameters. The equations of states are known exactly \cite{PhysRevX.6.041065,PhysRevLett.117.207201} by using the methods of the thermodynamic Bethe ansatz (TBA) \cite{KorepinBOOK,Takahashi72,Zamolodchikov1990} and its refinements \cite{String_charge,IQC17,Alba2017}. GHD, as it is currently developed, is then a Euler hydrodynamics based on these equations of state. Its power lies in part in the fact that the underlying TBA description is extremely universal, taking as input only few properties that are readily available for almost all integrable models. It is however a crucial question to understand the diffusive, Navier-Stokes corrections to Euler hydrodynamics (other types of corrections due to the lattice in free fermionic theories where studied in \cite{PhysRevB.96.220302}). In certain cases, spin and charge transport in quantum chains has been observed to show diffusive and other non-Eulerian behaviors \cite{PhysRevLett.103.216602,PhysRevE.96.020105,1801.07031,Ljubotina_nature,PhysRevB.90.155104,MKP17,PhysRevB.89.075139,1367-2630-19-3-033027,1711.11214,1712.09466}. { Diffusion also occurs in gases of hard rods \cite{Spohn1991,Lebowitz1967,1742-5468-2017-7-073210,Boldrighini1997} due to fluctuating accumulations of displacements, whence similar effects might be expected in soliton gases \cite{El2003,PhysRevLett.95.204101,El2010,PhysRevLett.120.045301,Medenjak17,1712.09466,1807.05000}
}.  Moreover it has been argued that what distinguishes interacting integrable models from free models should be diffusion \cite{1707.02159}. Is there then diffusion in integrable models { more generally}? If so, how universal is it, what form does it take? How does it modify the ballistic transport? 

 In this letter {we show that,  \emph{there is generically diffusive transport in interacting integrable models}} by providing an exact expression for the diffusion matrix.  We use this result to write an {\em exact and universal expression} for the Navier-Stokes term to the Euler-GHD hydrodynamic theory. We derive these expressions in the Lieb-Liniger (LL) model by performing a microscopic calculation of the Kubo formula \eqref{kubo}.  Our results agree with the known diffusion matrix in the hard rod gases \cite{Spohn1991,Boldrighini1997,1742-5468-2017-7-073210}, and we provide numerical checks for its validity in the anisotropic Heisenberg XXZ spin chain. Therefore we conjecture that our results are universal for every integrable models. We show that diffusive terms are responsible for positive entropy production, and we evaluate the diffusive, large-time corrections to the non-equilibrium currents in the partitioning protocol \cite{Spohn1977,1742-5468-2008-11-P11003,1751-8121-45-36-362001,PhysRevB.89.214308,PhysRevA.91.021603,1751-8121-48-9-095002,SciPostPhys.3.3.020,PhysRevB.90.161101,SciPostPhys.1.2.014,PhysRevE.96.012138,PhysRevB.95.045125,PhysRevB.90.155104,PhysRevX.7.021012,1742-5468-2016-6-064005,CDV18,1804.04476,1742-5468-2018-3-033104,PhysRevLett.120.176801}.

{\em Diffusion in hydrodynamics.---}
We first recall how diffusion is accounted for in linear fluctuating hydrodynamics \cite{Spohn1991,PhysRevLett.87.040601,OrtizdeZarate_book,1742-5468-2015-3-P03007,Spohn2016}. The general setting applies equally to conventional and generalised hydrodynamics. Its main objects are linear response and correlation functions of local charges and currents in homogeneous steady state. The static covariance matrix $C$ of conserved charges is $C_{ij} = \int_\R \dd x\,\bra \mathfrak{q}_i(x,0) \mathfrak{q}_j(0,0)\ket^{\rm c}$, where $\bra \mathfrak{q}_i(x,0)\mathfrak{q}_j(0,0)\ket^{\rm c} =\bra \mathfrak{q}_i(x,0)\mathfrak{q}_j(0,0)\ket - \bra \mathfrak{q}_i\ket \bra \mathfrak{q}_j\ket$.  By definition, if $\beta_i$ is the Lagrange parameter conjugate to the conserved charge $Q_i$ as in \eqref{GGE}, then $C_{ij} = -\p \bra \mathfrak{q}_i\ket/\p\beta_j$. The covariance matrix is therefore a property of stationary homogeneous states. An important quantity in non-equilibrium physics, part of Euler hydrodynamics, is the Drude weight, or Drude matrix $D_{ij}= \lim_{t\to\infty}\int_\R \dd x\, \bra \mathfrak{j}_i(x,t)\mathfrak{j}_j(0,0)\ket^{\rm c}$. If nonzero, it indicates the presence of ballistic transport. Beyond Euler, the effects of diffusion are encoded within the diffusion matrix $\mathfrak D$. Diffusion is not precluded by the presence of ballistic transport (nonzero Drude weight): it provides subleading corrections \cite{PhysRevB.83.035115,1707.02159,1702.08894,1806.04156}. In the conventional Kubo definition,
\beq\label{kubo}
	(\frak{D}C)_{ij} = \int_\R \dd t\,\lt(\int_\R \dd x\,\bra \mathfrak{j}_i(x,t)\mathfrak{j}_j(0,0)\ket^{\rm c} - D_{ij}\rt).
\eeq
evaluated on a generic stationary state. 
The diffusion matrix, evaluated at the space-time dependent local state \eqref{GGE}, fixes the Navier-Stokes term in \eqref{constitutive},
\beq\label{ADF}
	\mathfrak{D}_{ij} = -2{\mathcal F}_{ij}.
\eeq

Note that in homogeneous steady states, dynamical correlation functions $S_{ij}(k,t) = \int \dd x\,e^{-\ri kx}\bra \mathfrak{q}_i(x,t) \mathfrak{q}_j(0,0)\ket^{\rm c}$ take the form  $\exp\lt[-\ri k A t - \frc12k^2 \mathfrak{D} |t|\rt]C$, valid at large $t$, small $k$, where the ballistic propagation matrix is $A_{ij} = \p{\cal F}_i/\p \bra \mathfrak{q}\ket_j$. The term proportional to $kt$ represents ballistic transport, and that involving $k^2|t|$ the diffusive broadening of order $\sqrt{t}$ around the ballistic path. 

{\em Exact diffusion in GHD.---}  
In integrable models, stable quasi-particles exist whose scattering is elastic and factorised. As a consequence, a stationary homogeneous thermodynamic state (a GGE) in the thermodynamic limit can be fully characterised by a density of quasi-particles (microcanonical GGE \cite{PhysRevLett.110.257203}): the spectral density  $\rho_{\rm p}(\theta)$, with spectral parameter $\theta$ encoding both momentum and quasi-particle type. Each quasi-particle carries a quantity $h_i(\theta)$ of the charge $Q_i$,
\beq\label{qrho}
	\bra \mathfrak{q}_i\ket = \int \dd\theta\,h_i(\theta)\rho_{\rm p}(\theta)
\eeq
(the integral being implicitly accompanied by a sum over quasi-particle types in case there are many of them). For instance, in the repulsive LL model, there is a single type of quasi-particle and we choose $\theta$ to be the velocity, $h_0(\theta)=1$ for the actual particle density, $h_1(\theta) = p(\theta) = m\theta$ for the momentum and $h_2(\theta) = E(\theta) = m\theta^2/2$ for the energy. The interaction is fully encoded within the two-body scattering kernel $T(\theta,\alpha) = (2\pi \ri)^{-1} \,\dd \log S(\theta,\alpha)/\dd\theta$ \footnote{In integrable QFT, the differential scattering is usually denoted $\varphi(\theta,\alpha)= -\ri \,\dd \log S(\theta,\alpha)/\dd\theta$.}, where $S(\theta,\alpha)$ is the two-body scattering matrix. In the LL model, we have $T(\theta,\alpha) = c/(\pi( (\alpha - \theta) + c^2))$ where $c$ is the coupling constant between the bosonic particles. The TBA gives exact quasi-particle spectral densities in thermal states and GGEs \cite{Takahashi72,Zamolodchikov1990,PhysRevLett.115.157201,String_charge}, and in GHD, $\rho_{\rm p}(\theta; x,t)$ describes the local state \eqref{GGE}, via \eqref{qrho}.

At the Euler scale, the GHD equation is \cite{PhysRevX.6.041065,PhysRevLett.117.207201}
\beq\label{euler}
	\p_t \rho_{\rm p}(\theta; x,t) + \p_x (v^{\rm eff} (\theta; x,t)\rho_{\rm p}(\theta; x,t))=0.
\eeq
The {\em effective velocity} $v^{\rm eff}(\theta; x,t)$ represents the velocity of the quasi-particle of spectral parameter $\theta$, which is the group velocity \cite{PhysRevLett.113.187203} renormalised by the interactions with the other particles inside the local state at $x,t$. It solves the linear integral equation $p'(\theta)v^{\rm eff}(\theta) = E'(\theta) +(2\pi) \int{ \dd\alpha}{}\,\rho_{\rm p}(\alpha) T(\theta,\alpha)(v^{\rm eff}(\alpha) - v^{\rm eff}(\theta))$. The root of \eqref{euler} is the expression ${\cal F}_i = \int \dd\theta\,v^{\rm eff}(\theta)\rho_{\rm p}(\theta)h_i(\theta)$ for the Euler term of the constitutive relation \eqref{constitutive}. In fact, at the Euler scale, every local average $\bra \Or(x,t)\ket$ is evaluated within the GGE described by $\rho_{\rm p}(x,t;\theta)$, and indeed in a GGE the currents take the simple form $\bra \mathfrak{j}_i\ket = \int \dd\theta\,h_i(\theta)v^{\rm eff}(\theta)\rho_{\rm p}(\theta)$. Beyond the Euler scale, \eqref{qrho} stays true by definition but averages of other local observables have corrections that depend on the first derivative of the state variable, and in particular average currents get modified by the diffusion matrix, via \eqref{constitutive} and \eqref{ADF}.

The static covariance $C$, the Drude weight $D$ and the ballistic propagation matrix $A$ were evaluated exactly in \cite{SciPostPhys.3.6.039} in the full generality of Euler GHD. Three ingredients are involved: (i) the {\em occupation function} $n(\theta) = \rho_{\rm p}(\theta)/\rho_{\rm s}(\theta)$, where $\rho_{\rm s}(\theta) = \frac{p'(\theta)}{2\pi}+ \int \dd\alpha\,T(\theta,\alpha)\rho_{\rm p}(\alpha)$ is the state density; (ii) the {\em dressing} $h^{\rm dr}(\theta)$ of scalar spectral functions $h(\theta)$, a state-dependent modification of $h(\theta)$ expressed as the solution to the linear integral equation $h^{\rm dr}(\theta) = h(\theta) + \int  {\dd\alpha}{ }\,T(\alpha,\theta)n(\alpha) h^{\rm dr}(\alpha)$; and (iii) the {\em  statistical factor} $f(\theta)$, equal to $1-n(\theta)$ if the quasi-particle has fermionic statistics (such as in the LL model), $1+n(\theta)$ for bosonic statistics, $1$ for classical statistics and $n(\theta)$ for radiative modes (such as in classical field theory) \cite{1712.05687}.
\begin{figure}[t]
\includegraphics[width=0.36\textwidth]{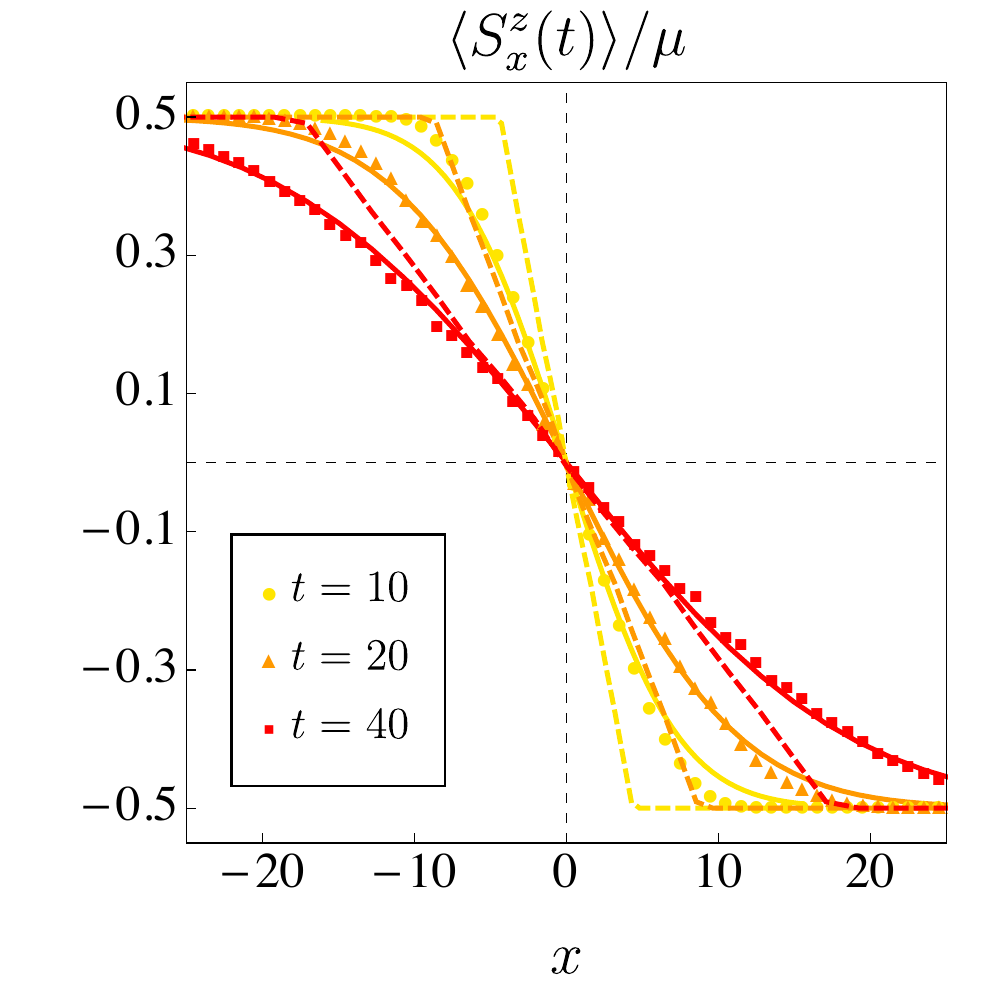}
  \caption{Plot of the magnetization profile $\langle S_x^z(t) \rangle/\mu$ at different times with the evolution given by an XXZ chain with anisotropy $\Delta=\cos(\pi/7)$ and with initial conditions given by the (small) domain wall initial state with density matrix given by $\hat{\rho}_0 \propto \prod_{x<0} e^{\mu S_x^z} \otimes \prod_{x>0} e^{ -\mu S_x^z}$ and $\mu =0.05$, namely $\langle S_x^z(t) \rangle = \text{Tr}[e^{ i H_{\text{XXZ}}t}  \hat{\rho}_0 e^{- i H_{\text{XXZ}}t} S_x^z]$. Continuous lines are the predictions of equation \eqref{eq:melting}, dashed lines are the predictions of Euler GHD  \eqref{euler} and dots are tDMRG numerical simulation reported in \cite{Ljubotina_nature}. }
  \label{fig:melting}
\end{figure}

The main result of this paper is a general, exact expression for the diffusion matrix in integrable models entering the Kubo formula \eqref{kubo} on a generic stationary state. We have derived it in the LL model based on natural conjectures on the matrix elements (form-factors) of conserved densities \footnote{The local densities $\mathfrak{q}_i$ are only fixed by the conserved charges $Q_i$ up to derivative terms. Technically, we fix this gauge ambiguity by fixing the singular structure of their matrix elements.}. We find the diffusion matrix
%
\begin{align}\label{diff}
  (\mathfrak{D} & C)_{ij}  =   \int \frc{\dd\theta\dd\alpha}{2}\,\rho_{\rm p}(\theta)f(\theta) \rho_{\rm p}(\alpha)f(\alpha)\,|v^{\rm eff}(\theta)-v^{\rm eff}(\alpha)| \no \\& \times 
\left(\mathfrak{A}_i(\theta,\alpha) - \mathfrak{A}_i(\alpha,\theta) \right)   \left( \mathfrak{A}_j(\theta,\alpha) -  \mathfrak{A}_j(\alpha,\theta) \right)
\end{align}
 where $  \mathfrak{A}_i(\theta,\alpha) =  h_i^{\rm dr}(\theta)  {T^{\rm dr}(\theta,\alpha)} ( \rho_{\rm s}(\theta))^{-1} $. The dressed differential scattering phase $T^{\rm dr}(\theta,\alpha)$ is the dressing of  $T(\theta,\alpha)$ as a vector field in its first argument $\theta$, solving $T^{\rm dr}(\theta,\alpha) = T(\theta,\alpha) + \int  {\dd\omega}{ }\,T(\theta,\omega)n(\omega) T^{\rm dr}(\omega,\alpha)$. Expression \eqref{diff} depends on the state via $\rho_{\rm p}$, $\rho_{\rm s}$, the effective velocity and the dressing operation. { \red Writing matrices in their dual integral-kernel form, $\mathfrak{D}_{ij} = (h_i,\mathfrak{D}h_j)=\int \dd\theta\dd\alpha\, h_i(\theta) \mathfrak{D}(\theta,\alpha)h_j(\alpha)$, and using matrix kernel multiplication $(A B)(\alpha,\theta) = \int d\gamma  A(\alpha, \gamma) B(\gamma, \theta)$, it is possible to extract an expression for the diffusion kernel {  $\mathfrak{D}=(1-nT)^{-1}\rho_{\rm s}\widetilde{\mathfrak{D}}\rho_{\rm s}^{-1}(1-nT)$ where $n$, $\rho_{\rm s}$ and $1$ are seen as diagonal integral kernels, with $   \rho_{\rm s}(\theta)^2\widetilde{{\mathfrak{D}}}(\theta,\alpha) = [w(\theta)\delta(\theta-\alpha) - W(\theta,\alpha)]$, where} $W(\theta,\alpha) = \rho_{\rm p}(\theta)f(\theta) \big[T^{\rm dr}(\theta,\alpha)\big]^2|v^{\rm eff}(\theta)-v^{\rm eff}(\alpha)|$ and $w(\theta) = \int \dd\alpha\,W(\alpha,\theta)$ (with the parametrisation choice such that $T$ is symmetric)}. We can now give a space- and time-dependence to the local GGE state $ \rho_{\text{p,s}}(\theta) \to  \rho_{\text{p,s}}(\theta;x,t) $ and the hydrodynamic equations \eqref{euler} are then modified by a Navier-Stokes term:
\begin{empheq}[box=\fbox]{equation}\label{ns}
	\p_t \rho_{\rm p} + \p_x (v^{\rm eff}\rho_{\rm p})=
	\frc12 \p_x \lt(\mathfrak{D} \p_x\rho_{\rm p} \rt)
  \end{empheq}
where $(\mathfrak{D} \partial_x\rho_{\rm p})(\theta; x,t ) = \int \dd\alpha \,\mathfrak{D}(\theta,\alpha; x,t) \partial_x\rho_{\rm p}(\alpha; x,t )$.
Although derived in the LL model, the expression \eqref{diff} is expressed in complete generality, as a function of the differential scattering kernel $T(\theta,\alpha)$ and the statistical factor $f(\theta)$. We conjecture that it applies to any integrable model, including Galilean and relativistic quantum and classical field theory, quantum chains, and classical gases \cite{Boldrighini1997,PhysRev.171.224,Spohn1982,Spohn1991}. For a gas of hard rods, where $T(\theta,\alpha)$ is a constant and $f(\theta)=1$, the diffusion operator is known exactly  \cite{Spohn1991,1742-5468-2017-7-073210,Boldrighini1997}, and we have verified that \eqref{diff} reproduces it correctly (see \cite{SM}).
In free (fermionic or bosonic) models, the differential scattering phase is zero, and thus no diffusion occurs. This confirms the proposition made in \cite{1707.02159}, and applies for instance to the infinite-coupling (Tonks-Girardeau) limit of the LL model. Expanding at large coupling $c$ \cite{2005_Brand_PRA_72}, the leading term of the diffusion matrix is in $1/c^2$, and we observe that it exactly agrees with the diffusion matrix for the hard rod gas. Often there is a choice of spectral parameter such that $T(\theta,\alpha) = T(\alpha,\theta)$, for instance when $\alpha$ is the velocity (rapidity), in most Galilean (relativistic) models. Symmetry of $T$ implies symmetry of $T^{\rm dr}$, simplifying \eqref{diff}. As a consequence, one has $\int \dd\theta\, p'(\theta) \mathfrak{D}(\theta,\alpha)=0$. This sum rule also follows from Galilean (relativistic) invariance, as then the current of mass (energy) is the momentum density. It is the defining property of a ``Markov operator" with respect to the measure $\dd p(\theta)$, extending the observation made for the hard rods \cite{Spohn1991,Boldrighini1997,1742-5468-2017-7-073210,Medenjak17}.

{\em Entropy production.---} A fundamental property of diffusion is that it causes an increase of the entropy of the local fluid cells. This is not in contradictions with unitary evolution of isolated systems: diffusion is indeed the passage of entropy from small-scale cells to large-scale structures. We therefore consider the total entropy of all fluid cells, $S = \int \dd x\,s(x)$. The entropy of integrable models takes different forms depending on the statistics of the quasi-particles. In classical particle models $s = -\int \dd\theta\, \rho_{\rm s} n \log n$, and in models with fermionic statistics $s = -\int \dd\theta \,\rho_{\rm s} (n \log n+(1-n)\log(1-n))$. In general, it is $s=\int \dd\theta\, \rho_{\rm s} g$ where $g$, seen as a function of $n$, satisfies $\p^2 g/\p n^2 = 1/(nf)$. We have shown (see the Supplemental Material) that if the operator $\mathfrak{D}C$ is positive (which is the case \eqref{diff}), the total fluid-cell entropy production is positive, and given by { $\p_t S = \int \dd x\, (\sigma,\mathfrak{D}C\, \sigma)/2$ with the choice of $\sigma$ such that $\sigma^{\rm dr} = \p_x n/(nf)$. }

{\em Sketch of proof.---} The derivation of \eqref{diff} in the LL model is based on the form-factors expansion of the dynamical current-current correlation function in the Kubo formula \eqref{kubo} into intermediate states of particles $\{\theta_{\rm p}^a\}$ and holes $\{\theta_{\rm h}^a\}$ microscopic excitation above the GGE stationary state $|\rho_{\rm p}\rangle$:
\beqa
	\lefteqn{\bra \frak{j}_i(x,t)\frak{j}_j(0,0)\ket^{\rm c}} &&\n &=&
	\sum_{n=1}^\infty \frac{1}{(n!)^2}\int \prod_{a=1}^n \lt[\dd\theta_{\rm p}^a \dd\theta_{\rm h}^a e^{\ri \lt( (k(\theta_{\rm p}^a)- k(\theta_{\rm h}^a)) x -(\ep(\theta_{\rm p}^a)-\ep(\theta_{\rm h}^a))t\rt)}\rt]\n
	&&\quad\times\,
	 \bra \rho_{\rm p}|\frak{j}_i|\{\theta_{\rm p}^\bullet, \theta_{\rm h}^\bullet\}\ket\bra \{\theta_{\rm p}^\bullet, \theta_{\rm h}^\bullet\}|\frak{j}_j|\rho_{\rm p}\ket
\eeqa
where $k(\theta)$ and $\ep(\theta)$ are appropriate momentum and energy function satisfying $\ep'(\theta) = v^{\rm eff}(\theta) k'(\theta)$. The current conservation equations \eqref{conslaw} imply the structure
\beq\label{struct}
	\bra \rho_{\rm p}|\frak{j}_i|\{\theta_{\rm p}^{\bullet},\theta_{\rm h}^{\bullet}\}\ket
	= \sum_a (\ep(\theta_{\rm p}^a)-\ep(\theta_{\rm h}^a))\,f_i(\{\theta_{\rm p}^{\bullet},\theta_{\rm h}^{\bullet}\})
\eeq
for the form-factors $\bra \rho_{\rm p}|\frak{j}_i|\{\theta_{\rm p}^{\bullet},\theta_{\rm h}^{\bullet}\}\ket$ on a non-trivial background (similar to finite-temperature form-factors \cite{Doyon2005,Doyon2007,1742-5468-2018-3-033102}). 
 According to general principles, $f_i$ has simple ``kinematic poles" at $\theta_{\rm p}^a = \theta_{\rm h}^b$ \cite{9789810202446,Bernard1993,PhysRevB.78.100403,Pozsgay2008,Doyon2005}. 
The integral over $x$ in \eqref{kubo} gives $\delta(\sum_a (k(\theta_{\rm p}^a)-k(\theta_{\rm h}^a)))$. At $n=1$, this imposes equality of particle and hole momentum. Interpreting a hole excitation as an outgoing particle, this equality is interpreted as ballistic propagation.  The exact value of the one particle-hole pair matrix element at equal particle and hole momenta is known \cite{1742-5468-2018-3-033102,PhysRevLett.120.217206}, and gives  the Drude weight \cite{IN_Hubbard} (see also \cite{SM}). At $n=2$, with two particles and two holes, the integral over time in \eqref{kubo} gives the extra factor $\delta(\sum_a (\ep(\theta_{\rm p}^a)-\ep(\theta_{\rm h}^a)))$. The two delta functions are now simultaneous conservation of momentum and energy in a two-body scattering process. In 1+1 dimension, this imposes equality of the sets of incoming and outgoing momenta. The combination of the ``energy-conservation" factor $\sum_a (\ep(\theta_{\rm p}^a)-\ep(\theta_{\rm h}^a))$ in \eqref{struct} and equality of the sets of incoming (particle) and outgoing (hole) momenta imply that it is the kinematic poles of the functions $f_i$ that provide a nonzero result at $n=2$. Evaluating this using known kinematic residues, we obtain \eqref{diff}. For $n>2$, the energy-conservation factor gives zero against $\delta(\sum_a (\ep(\theta_{\rm p}^a)-\ep(\theta_{\rm h}^a)))$, as the set of incoming and outgoing momenta are no longer conserved with more than 2 particles. See \cite{SM} for full details of the computation.
%
%

{\em Discussion.---} The proof we sketched above has a natural interpretation: \emph{the contribution to diffusion comes from two-body scattering events between quasiparticle excitations above the local GGE state}. Like the effective propagation velocity, the scattering amplitude is renormalised by the local fluid state, thus the diffusion matrix \eqref{diff} involves the dressed scattering kernel $T^{\rm dr}(\theta,\alpha)$. Although the ballistic matrix $A$ is diagonal in the quasiparticles momenta -- quasiparticles propagate ballistically in a homogeneous GGE -- the diffusion matrix $\mathfrak{D}$ is not diagonal, as to first order beyond the Euler scale elastic two-body scattering events are important, reminiscent of how two-body scattering terms in Boltzmann's equation lead to diffusion in standard hydrodynamics.  Higher-body scattering processes are expected to take place at sub-diffusive orders.

{\em Numerical evaluations.---}
It is instructive to consider near-homogeneous and stationary situations, $\rho_{\rm p}(\theta;x,t)\sim\rho_{\rm p}^{\rm sta}(\theta)$ and $\p_x \rho_{\rm p} \ll 1 $. In terms of the occupation function $n(\theta; x, t )$ equation \eqref{ns} becomes
{\beq\label{eq:melting}
	\p_tn + v^{\rm eff}\p_x n = \frc12 \widetilde{\mathfrak{D}} \p_x^2 n + O( (\p_x n)^2 )
\eeq
 where the operator $\widetilde{\mathfrak{D}}(\theta,\alpha)$ is evaluated on the stationary state in the linear approximation}. Eq. \eqref{eq:melting} represents a diffusive spreading correcting the ballistic propagation. Let us consider the {\em partitioning protocol} for the construction of non-equilibrium steady states, where two semi-infinite baths, initially independently thermalised, are then connected and let to evolve unitarily, see Fig. \ref{fig:melting}. The solution at the Euler scale in integrable models is a continuum of contact singularities, one for each value of $\theta$ \cite{PhysRevX.6.041065,Doyon2018}. These singularities are affected by diffusive spreading, which, upon integration over $\theta$, gives rise to $1/\sqrt{t}$ corrections to local observables.  We compared with exact tDMRG numerics in the XXZ Heisenberg chain {\red $H_{\text{XXZ}}= \sum_x \big(S^+_x S^-_{x+1} + S^-_x S^+_{x+1}   + \Delta/2 S^z_x S^z_{x+1}\big)$} (with $S_x^\alpha$, $\alpha=+,-,z$ the spin-$1/2$ operator at position $x$) in the gapless regime $|\Delta|<1$. We find that corrections due to the diffusive term dramatically {\em improve} the Euler-scale predictions \cite{PhysRevLett.117.207201}, see FIG.~\ref{fig:melting}.
We also tested our formula \eqref{diff}, in a XXZ spin chain at thermal equilibrium with infinite temperature and zero total magnetisation, against numerical tDMRG data \cite{PhysRevB.89.075139,1367-2630-19-3-033027}. We considered $\Delta = \cos \pi /\ell$ for integer $\ell$, where the number of quasi-particle types is finite, given by $\ell$ \footnote{Notice that in the limit $n,m \to \infty$  with $\Delta = \cos \pi n/m$, the spin matrix elements of $\mathfrak{D}$ diverge due to the infinite number of quasiparticle, in accord with the recent results for the spin diffusion constant reported in \cite{1806.03288}.}. For the spin diffusion constant $(\mathfrak{D}C)_{S^zS^z}$, from  formula \eqref{diff} we obtained $0.137$, $0.281$, $0.744$ for $\ell=3,\,4,\,7$ resp., agreeing with $0.14\pm0.01$, $0.24\pm0.05$, $0.72\pm0.05$ for the integrated spin current-current correlator \eqref{kubo} obtained with tDMRG. 

{\em Conclusion.---} We derived large-scale hydrodynamics equations accounting for diffusive effects in integrable models. These equations complete the Euler-scale hydrodynamic approach introduced originally in \cite{PhysRevX.6.041065,PhysRevLett.117.207201} and allow to access shorter time and length scales. We checked our results by reproducing analytically the known expression in classical hard rod gases and by numerical comparisons with tDMRG numerical data for an XXZ spin chain, finding excellent agreement. Extension to different models such as the gapped XXZ chain and the Fermi-Hubbard model \cite{PhysRevB.90.155104}, the dynamics of integrable spin chains with weak coupling to external environment \cite{1367-2630-12-4-043001,Lange2017,PhysRevB.97.165138},  the effects of an external trapping potential on diffusive phenomena and thermalisation \cite{SciPostPhys.2.2.014,PhysRevLett.120.164101,1711.00873} and super-diffusive transport in the presence of isotropic interactions \cite{1742-5468-2015-3-P03007,Ljubotina_nature,1806.03288} are under current investigation.

{\em Acknowledgments.---} 
The authors kindly  acknowledge Christoph Karrasch and Marko Ljubotina for providing tDMRG data and Enej Ilievski, Tomaz Prosen, Adrea De Luca and Koenraad Schalm for stimulating discussions and valuable comments.
J.D.N. acknowledges support from LabEx ENS-ICFP:ANR-10-LABX-0010/ANR-10-IDEX-0001-02 PSL*. D.B. acknowledge support from ANR contract ANR-14-CE25-0003. B.D. thanks the \'Ecole Normale Sup\'erieure de Paris for an invited professorship from February 19th to March 20th 2018, where a large part of this work was carried out, and he also thanks the Centre for Non-Equilibrium Science (CNES) and the Thomas Young Centre (TYC). All the authors acknowledge hospitality and funding from the Erwin Schr\"odinger Institute in Vienna (Austria) during the program ``Quantum Paths" from April 9th to June 8th 2018 and the GGI in Florence (Italy) during the program ``Entanglement in Quantum Systems".

\bibliography{references}

\appendix
\onecolumngrid

\vspace{1cm}
\begin{center}
\textbf{{\Large Supplemental Material}}
\end{center}

%

\section*{Entropy production}

In this section we show that, from the diffusion term obtained, total entropy production is positive. For this purpose, we consider generic statistics, represented by the function $f$ as in the main text. Here and below, we see $f$ as a function of $n$, as its full dependence on the spectral parameter and space-time is via the occupation function $n(\theta;x,t)$. Recall that for fermions $f=1-n$ (such as in the Lieb-Liniger model and many other integrable quantum models), for bosons $f=1+n$ (such as in the quantum Klein-Gordon model), for classical particles $f=1$ (such as in the free classical gas or the hard rod gas) and for classical radiative modes $f=n$ (such as in the classical Klein-Gordon field or the classical sinh-Gordon field); other statistics are also possible. The local state described by the distribution $\rho_{\text{p}}(\theta;x,t)$ has finite entropy given by the formula
\begin{equation}
s(x,t) =  -\int \dd\theta\, \rho_{\text{s}}(\theta;x,t)g(n(\theta;x,t))
\end{equation}
where $g(n)$ satisfies the equation
\beq
	\frc{\p^2 g}{\p n^2} = \frc1{nf}.
\eeq
This function is specified up to terms of zeroth and first power in $n$, which do not affect entropy production. One can show that in the Fermionic case the Yang-Yang formula is recovered,
\begin{equation}
s =  -\int \dd\theta\, \rho_{\text{s}}\left( n\log n + (1-n)\log( 1- n) \right),
\end{equation}
in the bosonic case that of Zamolodchikov is obtained
\beq
s =  -\int \dd\theta\, \rho_{\text{s}}\left( n\log n - (1+n)\log( 1+ n) \right),
\eeq
in the case of classical particles one finds the usual classical entropy,
\beq
s =  -\int \dd\theta\, \rho_{\text{s}}n\log n,
\eeq
and the correct classical field entropy is found in the case of radiative modes,
\beq
s = \int \dd\theta\, \rho_{\text{s}}\log n.
\eeq
Here and below, we omit the explicit space-time and spectral parameter dependence for lightness of notation.

The total entropy of the system is $S = \int \dd x \,s$.  At Euler scale the local entropy $s$ satisfies a continuity equation, namely
\beq
	\p_t s + \p_x j_s = 0
\eeq
with
\beq
	j_s (x,t)= -\int \dd\theta\,v^{\rm eff}\rho_{\rm s} g,
\eeq
and therefore the total entropy is preserved, $\p_t S = 0$. We now introduce the Navier-Stokes terms and we show that the total entropy then is not constant. For simplicity we assume a parametrisation has been chosen such that $T$ to be symmetric, but the proof can easily be written for general parametrisation.

According to Eq. (14) in the main text, we have 
\begin{equation}\label{eq:1}
\p_t \rho_{\rm p} + \p_x (v^{\rm eff}\rho_{\rm p})=
	 \p_x  \mathcal{N}, \qquad   \mathcal{N} = \frc12 \mathfrak{D} \partial_x \rho_{\rm p},
\end{equation}
where, as is clear from the explicit formula in the main text, we see $\p_x \rho_{\rm p}$ as a vector in the space of spectral functions, and $\mathfrak{D}$ as an integral operator acting on this space. Let us define the conjugate $\widetilde{\mathfrak{D}}$ as in the main text by
\beq
	\mathfrak{D} = (1-nT)^{-1} \rho_{\rm s}\widetilde{\mathfrak{D}}\rho_{\rm s}^{-1} (1-nT),
\eeq
where $T$ is the integral operator with kernel $T(\theta,\alpha)$, and $n$ on is seen as a diagonal integral operator with diagonal elements $n(\theta)$. Note in particular that
\beq\label{eq:DC}
	\mathfrak{D}C = (1-nT)^{-1} \rho_{\rm s}\widetilde{\mathfrak{D}}n f(n)(1-Tn)^{-1},
\eeq
and also that
\beq\label{eq:hdr}
	h^{\rm dr} = (1-Tn)^{-1} h
\eeq
for scalar spectral functions $h$.  
First, we show that
\beq\label{eq:3}
	(1-nT){\cal N} = \frc12 \rho_{\rm s}\widetilde{\mathfrak{D}} \p_x n.
\eeq
Indeed, we have
\beq
	(1-nT){\cal N} = \frc12 \rho_{\rm s}\widetilde{\mathfrak{D}}\rho_{\rm s}^{-1}(1-nT)\p_x\rho_{\rm p}.
\eeq
Noting that $T\rho_{\rm p} = \rho_{\rm s}-p'/2\pi$, we find
\beq
	(1-nT)\p_x\rho_{\rm p} = \p_x \rho_{\rm p} -n\p_x (T\rho_{\rm p})
	= \p_x \rho_{\rm p}-n\p_x \rho_{\rm s}
	= \p_x \rho_{\rm p} -n\p_x \rho_{\rm p}/n
	= \rho_{\rm p}\p_x \log n.
\eeq
which shows \eqref{eq:3}. Second, we show that
\beq\label{eq:2}
	\p_tn + v^{\rm eff}\p_x n = \frc1{\rho_{\rm s}}(1-nT)\p_x {\cal N}.
\eeq
For this, we recall that, for any parameter $\mu$ on which $n$ may depend,
\beq
	\p_\mu \Big(n(1-Tn)^{-1}\Big) = (1-nT)^{-1} \p_\mu n (1-Tn)^{-1}.
\eeq
Therefore
\beq
	2\pi \p_t \rho_{\rm p} = \p_t \Big(n(1-Tn)^{-1}p'\Big) = 
	(1-nT)^{-1} \p_t n (1-Tn)^{-1} p',
\eeq
and similarly for $2\pi \p_x \rho_{\rm p}$, so we have from \eqref{eq:1}
\beq
	(1-nT)^{-1} \p_t n (p')^{\rm dr} + (1-nT)^{-1}
	\p_x n (E')^{\rm dr} = 2\pi\p_x{\cal N},
\eeq
and therefore
\beq
	\p_t n + v^{\rm eff} \p_x n = \frc{2\pi}{(p')^{\rm dr}}(1-nT)\p_x{\cal N}.
\eeq
which shows \eqref{eq:2}.

We now consider the functional
\beq
	s = -\int \dd\theta\,\rho_{\rm p}F(n),
\eeq
where $nF(n) = g(n)$. Extracting the terms involving ${\cal N}$ and using \eqref{eq:3} and \eqref{eq:2}, we have
\beqa
	\lefteqn{\p_t s + \p_x j_{s}} && \n &=& -\int \dd\theta\,\big(
	\p_x {\cal N}F(n) + \rho_{\rm p}F'(n)\times
	\frc1{\rho_{\rm s}}(1-nT)\p_x {\cal N}
	\big)\n
	&=& -\int \dd\theta\,\big(
	\p_x {\cal N}F(n) + n F'(n)(1-nT)\p_x {\cal N}
	\big)\n
	&=& -\p_x  j'_s + \int \dd\theta\,\big(
	{\cal N}F'(n)\p_x n + (F'(n)+nF''(n))(\p_x n) (1-nT) {\cal N} -
	n F'(n) (\p_x n T) {\cal N}
	\big)\n
	&=& -\p_x  j'_s + \int \dd\theta\,
	(2F'(n)+nF''(n))\p_x n (1-nT){\cal N} \n
	&=&-\p_x  j'_s  + \frc12\int \dd\theta\,
	(2F'(n)+nF''(n))\p_x n\, (\rho_{\rm s}\widetilde{\mathfrak{D}}\rho_{\rm s}^{-1})\rho_{\rm s}\,\p_x n.
\eeqa
where
\begin{equation}
  j'_s = \mathcal{N} F(n) + n F'(n) (1-n T)\mathcal{N}.
\end{equation}
Now we note that $2F'(n) + nF''(n) = g''(n) = 1/(nf(n))$, and we find 
\beq
	\p_t  s + \p_x (j_{s} +j'_s)=  \frc12\int \dd\theta \int \dd\alpha\,
	\frc{\p_x n(\theta)}{n(\theta)f(n)}\, \rho_{\rm s}(\theta)\widetilde{\mathfrak{D}}(\theta,\alpha)  {\p_x n(\alpha)}{ }.
\eeq
Using then eq. \eqref{eq:DC} this completes the statement in the main text. 

\end{document}